\documentstyle[12pt,cite,epsfig]{article}
\def\RI{$R_{1D}\ $}
\def\pb{$Pb$$-$$Pb$\ }
\def\fileversion{2.1}
\def\filedate{95/01/26}
\newlength{\dinwidth}
\newlength{\dinmargin}
\begin{document}
\begin{titlepage}

.

\begin{flushright}
\vspace{2cm}

{\bf TAUP 3020/17}
\end{flushright}
\vspace{2cm}

\begin{center}
{\large\bf  The Bose-Einstein Correlations\\
\vspace{1.5mm}
and the  strong coupling constant at low energies}\\
\vspace{15mm}

{\large\bf  Gideon Alexander$^{a,}$\footnote{Email: gideona@post.tau.ac.il}
and Boris Blok$^{b,}$\footnote{Email: blok@ph.technion.ac.il}}\\
\end{center}
\vspace{2mm}

\centering{\large{\it  a) School of Physics and Astronomy,
Raymond and Beverly Sackler Faculty of Exact Sciences,
Tel-Aviv University, 69978 Tel-Aviv, Israel\\
b) Department of Physics, Technion, Israel Institute of Technology,
32000 Haifa, Israel}}
\vspace{10mm}

\begin{abstract}
It is shown that $\alpha_s(E)$, the strong coupling constant,
can be determined in the non-perturbative
regime from Bose-Einstein correlations (BEC).
The obtained $\alpha_s(E)$ is in agreement with
the prescriptions dealt with in the
Analytic Perturbative Theory approach.
It also extrapolates smoothly to the standard
perturbative $\alpha_s(E)$
at higher energies. Our results indicate that
BEC dimension can be considered as an alternative
approach to the short range measure between hadrons.
\end{abstract}
\vspace{5mm}

\begin{center}
{\large \today }
\end{center}
\end{titlepage}

%\begin{flushleft}
\setcounter{footnote}{0}

\section{Introduction}
\label{sec1}
\hspace{6mm}
In recent years Bose-Einstein (BEC)  and Fermi-Dirac (FDC)
correlations have been extensively studied mainly with identical pion pairs
produced in lepton-lepton
and hadron-hadron reactions, as well as in heavy ion (AA)
collisions. In the one dimension (1D) correlation analyzes
of pion and hadron pairs it was found
that the resulting $R_{1D}$ dimension depends on the particle mass
and found to be proportional to 1/$\sqrt{m}$ where $M$ is
the mass of the correlated particles
(See e.g. Ref. \cite{review}).
It has been further shown \cite{genia}  that this $R_{1D}(m)$ behavior
can be described
in terms of the Heisenberg uncertainty relations
and from a  general QCD potential considerations.
\vspace{2mm}

\noindent
The  two identical particle correlation effect can be
measured in terms of the correlation function
\begin{equation}
C(p_1,p_2)\ =\ \frac{\rho(p_1,p_2)}{\rho_0(p_1,p_2)}\ ,
\label{bec}
\end{equation}
where $p_1$ and $p_2$ are the 4-momenta of the two correlated
hadrons and $\rho(p_1,p_2)$
is the two particle density function. The $\rho_0(p_1,p_2)$ stands for
the two
particle density function in the absence of the a BEC  (or FDC) effect.
This
$\rho_0$ is often referred to as the reference sample against
which the correlation effect is compared to.
The BEC and FDC analyzes had often different experimental
backgrounds and have chosen various types of reference
$\rho_0(p_1,p_2)$ samples. Thus one has to take this situation
in account when judging the correlation results
in terms of energy and/or mass dependence.
\vspace{2mm}

\noindent
The function frequently used in
the BEC and the FDC studies
%a
the evaluation of the $R_{1D}$ are:
\begin{equation}
C(Q)\ =\ 1+\lambda e^{-Q^2R^2_{1D}}\ {\rm{for\ bosons}}\
{\rm{and}} \ \ \ C(Q)\ =\ 1-\lambda e^{-Q^2R^2_{1D}}\  {\rm{for\ fermions}}
\label{gauss}
\end{equation}
\noindent
These are the  Goldhaber parametrizations \cite{goldhaber} proposed
for a static
Gaussian particle source in the plane-wave approach
which assumes for the particles emitter a
spherical volume with a radial Gaussian distribution.
The $\lambda$ factor, also known as the chaoticity
parameter, lies within  the range of 0 to 1.
Due to the fact
that the major correlation experiments were carried out
with identical bosons we will here focus our discussion
no the Bose-Einstein correlations.

\begin{figure}[ht]
\centering{\psfig{file=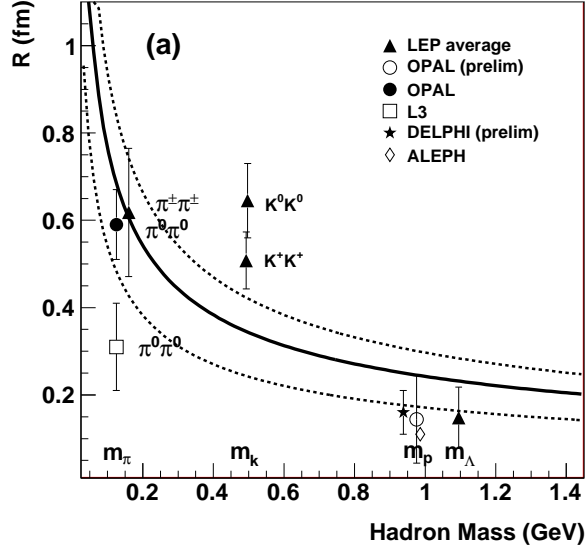,height=9.0cm,
bbllx=589pt,bblly=16pt,bburx=27pt,bbury=621pt,clip=}}
\caption{\small A compilation of \RI versus
the hadron mass obtained from BEC and FDC analyzes
of the $Z^0$ hadronic decays of the
LEP experiments \cite{review,delphi} taken from
Ref. \cite{reinherz}.
The solid
line and the dotted lines represent respectively
Eq. (\ref{formula}) with $\Delta t$=$10^{-24}$ sec
and $\Delta t$=$(1\pm 0.5)\times10^{-24}$ sec
.}
\label{z0decay}
\end{figure}
\begin{figure}[h]
\centering{\psfig{file=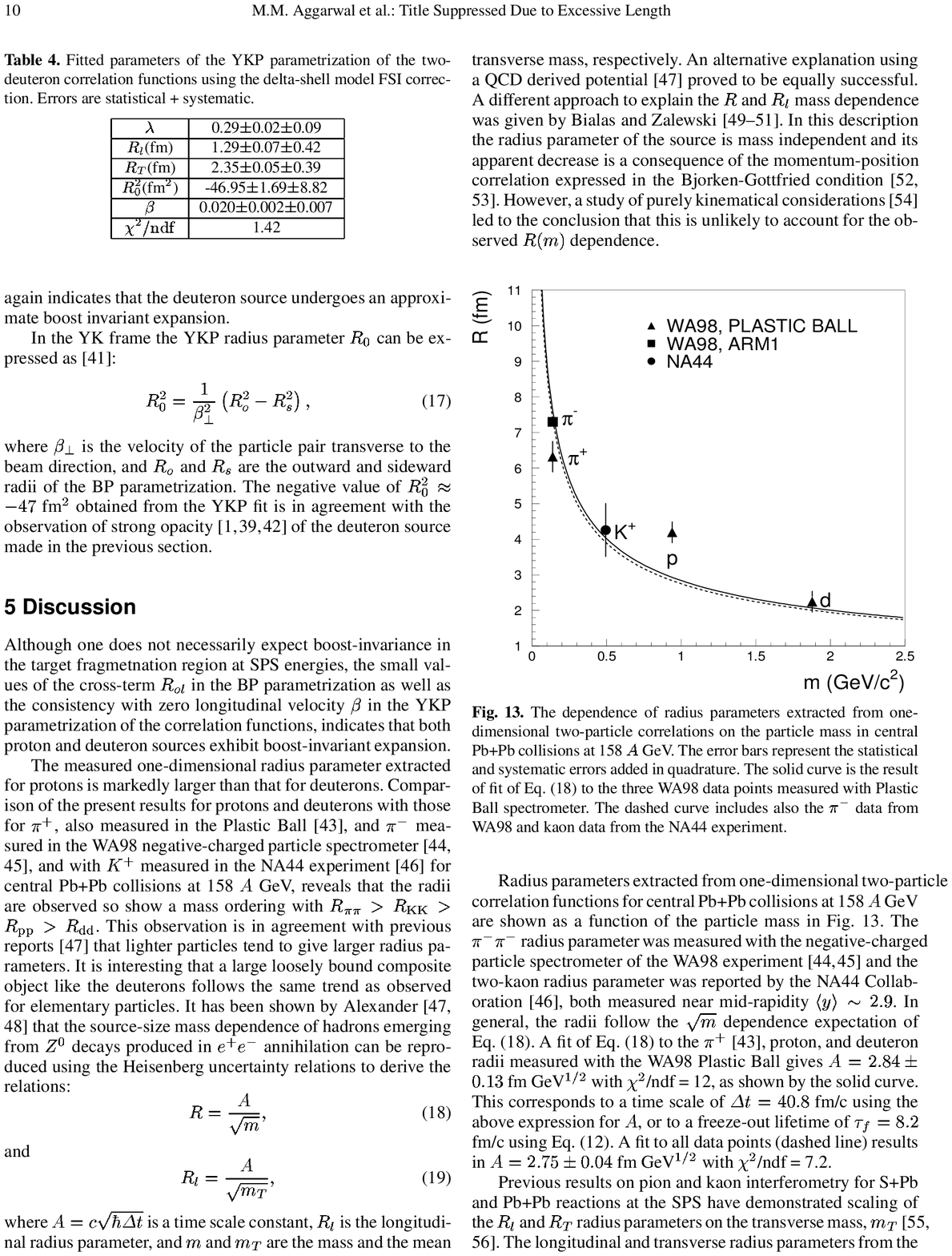,height=6.8cm,
bbllx=298pt,bblly=390pt,bburx=559pt,bbury=626pt,clip=}}
\caption{
BEC analyzes of hadron pairs
emerging from central \pb collisions at 158/A GeV \cite{wa98}.
The continuous line represents a fit of
Eq. (\ref{formula}) to the data of the Plastic Ball
detector. The result of the fit where the WA98 data and
the Kaon-pair \RI value from the NA44 collaboration were also
included
is shown by the dashed line.
}
\label{wa98mass}
\end{figure}
\vspace{2mm}

\noindent
It has been noticed already
some two decades ago
that the $R_{1D}$ extracted from
BEC and FDC analyzes of hadron pairs produced
in the decay of the $Z^0$ gauge boson
suggested a mass dependence  roughly proportional
to 1/$\sqrt{m}$ \cite{genia}. This is illustrated in
Fig. \ref{z0decay} taken from reference
\cite{reinherz} where a compilation of the
$R_{1D}$ results was obtained from the $Z^0$ hadronic decay
experiments at LEP, The difference between
the \RI value at the pion mass
to those of the proton and $\Lambda$ baryons
is indeed impressive. However presently no
significant difference is seen between the \RI
of the pions and the K-mesons produced in the $Z^0$ decay.
Thus this dimension data, deduced
from the $e^+e^-$ prompt interactions, cannot
serve for a precise expression for the \RI dependence on energy.
For that reason we
% have here
 utilize here the BEC dimension
results obtained in \pb collisions experiments \cite{wa98}.
\vspace{2mm}

\noindent

\noindent
In this letter we show that the BEC can serve in the
evaluation of the
strong coupling constant $\alpha_s$
at the non-perturbative region of $E<1$ GeV.
The resulting coupling constant is shown to be
in good qualitative agreement with the one obtained
from solving
the Bethe-Saltpeter equation to determine the
effective potential of the quarkonium
which in turn is consistent with the $\alpha_s$
deduced from an
Analytic Perturbative Theory (APT) prescription.

\noindent
In section 2 we discuss the mass dependence of the BEC source scale
and in section 3 we derive an analytic formula relating
the strong coupling constant $\alpha_s$ and
the BEC source radius.  Finally in section 4 we present numerical
values for $\alpha_s$ and show that the BEC derived source
dimension corresponds to a strong overlapping hadrons
where the BEC source radius is of the order
of the distance between them.

\section{The mass dependence of the BEC dimension $R_{1D}$}
\noindent
Since the maximum of the BEC enhancement of
two identical bosons of mass $m$
occurs when $Q\rightarrow 0$,  the three vector
momentum difference of the bosons approaches zero. Thus we can
link the BEC effect to the
Heisenberg uncertainty principle \cite{genia}, namely
\begin{equation}
\Delta p\Delta r\ =\ 2\mu vr\ =\ mvr\ =\ \hbar c\ ,
\label{uncertainty}
\end{equation}
where $\mu$ is the reduced mass of the di-hadron system and $r=\Delta r$
is the distance between them.  Here we use for
$\Delta p$ the GeV unit while $r$ is given
in fm units so that  $\hbar c$\ =\ 0.197 GeV fm.
Thus one obtains:
\begin{equation}
r\ =\ \frac{\hbar c}{mv}\ =\ \frac{\hbar c}{p}\ .
\label{ratio}
\end{equation}
Simultaneously we also apply the uncertainty relation expressed in terms
of time and energy
\begin{equation}
\Delta E \Delta t\ =\ \frac{p^2}{m}\Delta t\ =\ \hbar\ ,
\label{uncertainty2}
\end{equation}
where the energy and $\Delta t$ are given respectively in
GeV and seconds.
Thus one has
\begin{equation}
p\ =\ \sqrt{\hbar m/\Delta t}\ .
\end{equation}
Inserting this expression for $p$ into Eq. (\ref{ratio}) one finally obtains
\begin{equation}
r(m)\ =\ \frac{hc}{\sqrt{m}}\sqrt{\frac{\Delta t}{h}}     \ =\
 \frac{c\sqrt{\hbar \Delta t}}{\sqrt{m}}\ .
\label{formula}
\end{equation}
\vspace{2mm}
Comparing values of $r$ in Eq. \ref{formula} and experimental data for
\RI (see Fig. \ref{z0decay,
rpp_fig}) we are lead to identify $r$ with \RI.
\noindent

%%%%%%%%%%%%%%%%%%%%%%%%%%%%%%%%%%%%%%%%%%

As it was mensioned above the \RI values, deduced from the
BEC and FDC analyzes of the $Z^0$ hadronic decays
are shown in Fig. \ref{z0decay}.
These results provided the first
clue that the \RI may depend on the mass of the two identical
correlated particles \cite{genia}.
As can be seen, the measured $R_{1D}$ values
of the pion pairs are located
at $\sim 0.6$ fm with the exception of one $\pi^0\pi^0$ result
where its $R_{1D}$ value
lies significantly lower. The \RI Kaon pairs values
are seen to be near to those of the  charged pions.
Impressive however are the
\RI values obtained from the  $\Lambda$ hyperon and proton
baryon pairs
which lie close together in the vicinity of
0.15 fm.  The solid line in the figure was
calculated from
Eq. (\ref{formula}) with $\Delta t=10^{-24}$ sec representing the
strong interactions time scale. The dashed lines are derived from
Eq. (\ref{formula}) setting  $\Delta t=(1\pm 0.5)\times 10^{-24}$ sec
in order to illustrate the sensitivity of Eq. (\ref{formula}) in
its ability to estimate the energy dependence of \RI.
An alternative way to extract
\RI dependence on the energy
is to use
the BEC results of the boson pairs
produced in \pb collisions.
\vspace{2mm}

\begin{figure}[ht]
\centering{\psfig{file=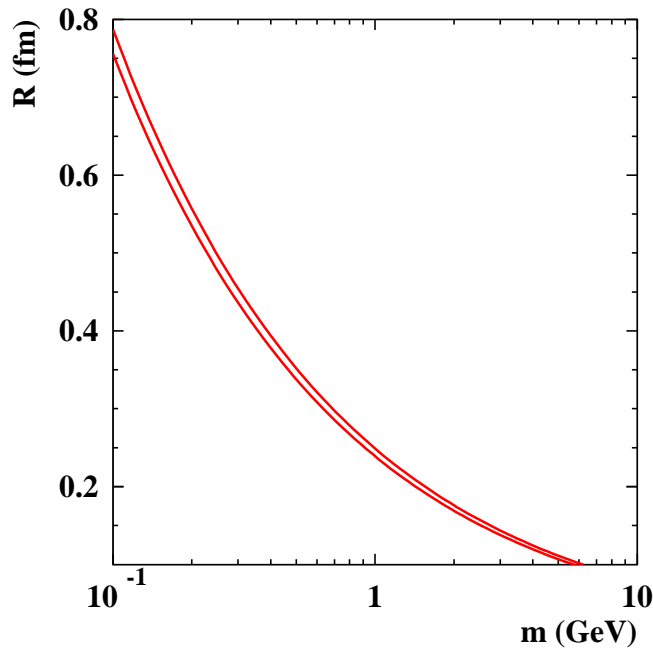,height=9.0cm,
bbllx=36pt,bblly=143pt,bburx=538pt,bbury=676pt,clip=}}
\caption{\small The $\pm$1 s.d. \RI band versus the mass energy
of the boson pairs produced at a time delay of $\Delta t= 10^{-24}$ sec
as deduced  from BEC of hadron pairs produced in \pb
collisions at 158 GeV/A \cite{wa98} with a time delay of
$\Delta t = (1.28\pm 0.04)\times10^{-22}$ sec (see text)..
}
\label{rpp_fig}
\end{figure}

\noindent
A clear evidence for the dependence of \RI on the mass of the BEC
boson pairs
is seen in Fig. \ref{wa98mass}
%%%%%%%%%%%%%%%%%
that
was  obtained by the WA98 collaboration
\cite{wa98}.
In  Figure \ref{wa98mass} are shown
%%%%%%%%%%%%%%%%%%%%%%
%are plotted
the BEC dimension deduced from identical correlated boson pairs,
including the deutron pairs,
produced in \pb collisions at the nucleon-nucleon center of mass
energy of 158 GeV/A. As can be seen,
apart from the proton pair result, the \RI dependence
on the mass value is very well described by
$A/\sqrt{m}$,
with the fitted value of $A=(2.75 \pm 0.04)$ fm GeV$^{1/2}$.
According to Eq. (\ref{formula}) one finds that
$A=c\sqrt{\hbar{\Delta t}}$
so that in the \pb collisions case
$\Delta t = (1.28\pm 0.04)\times10^{-22}$ sec.
Taking for prompt $pp$
collision the representing strong interaction value of
$\Delta t= 10^{-24}$ sec one obtains for \RI versus the
mass, in GeV units, the relation
\begin{equation}
R_{1D}\ =\ \frac{0.244\pm 0.005}{\sqrt{m(\rm{GeV})}}\ \ \rm{fm}\ ,
\label{rpp}
\end{equation}
which is shown by a $\pm$1 s.d. band in Fig. \ref{rpp_fig}
which is consistent with the LEP result.
%We further
%equate the distance $r$
%between hadrons to the source radius $R_{1D}$.

\section{The strong coupling constant and the BEC}
The short range interactions between two hadrons can be
described in terms of the  constituent quark model.
This idea dates back to \cite{Liberman} (see also \cite{FS1})
and was applied to the BEC in \cite{genia}.
Namely, the short range interaction between hadrons can be described by means of the
quark-quark interaction potential \cite{cornell}
\begin{equation}
V(r)=-(4/3)\alpha_s\hbar c/r+\kappa r\ .
\end{equation}
The coupling constant $\alpha_s$ is usually taken
as a parameter to be fitted, while the constant
$\kappa$, that corresponds to the
confinement part of the interaction,
is of order of 0.9 GeV/fm \cite{mendel},
while  $r$ is the distance between the two hadrons.

\noindent
We now make use of the virial theorem for the two hadron
system, which has the form
\begin{equation}
\left<2T\right >=<\vec r\cdot\vec  \nabla V(r)>\ ,
\end{equation}
where $\left<T\right >$ is the average kinetic energy of the
hadrons.
Taking into account the spherical symmetry of the potential and
the  uncertainty relations discussed above,
one obtains
\begin{equation}
 \frac{(\hbar c)^2}{m}=r^3\frac{dV}{dr}\ .
\end{equation}
This yields straightaway an expression for the strong coupling
constant, namely
\begin{equation}
r^3\left  (\kappa +\frac{4}{3}\frac{\alpha_s \hbar c}{r^2}\right
)-\frac{(\hbar c)^2}{m}=0\ ,
\label{eqvr}
\end{equation}
from which one has that  $\alpha_s$ is equal to
\begin{equation}
\alpha_s\ =\ \frac{3}{4}\frac{(\hbar c)^2 -r^3\kappa m}{mr\hbar c}\ .
\label{eqvirial}
\end{equation}

\noindent
Inserting $hc=0.1973$ GeV fm one obtains
\begin{equation}
\alpha_s=\frac{1.267(0.1168-3r^3km)}{mr}\ .
\label{sof}
\end{equation}
To evaluate $\alpha_s$ we use for the parameter $\kappa$ the
value of 0.18 GeV$^{2}/ 0.9\ $GeV/fm \cite{mendel} corresponding to
the meson Regge trajectory.
The variable $r$ and its mass dependence are taken to be
identical to the \RI dimension given by Eq. (\ref{rpp}) which was
determined from the analyzes
of the BEC and
FDC  deduced from identical hadron
pairs (see also  \cite{genia}).

\section{Conclusions.}
\begin{figure}[ht]
\centering{\psfig{file=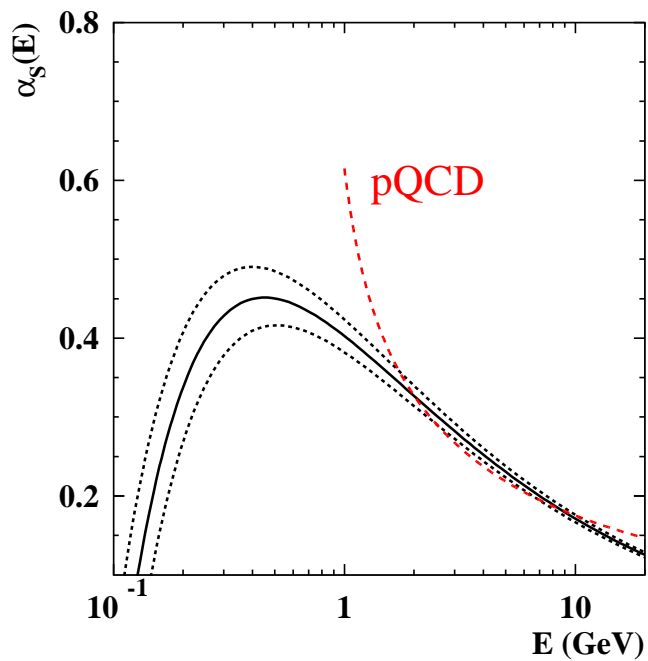,height=9.0cm,
bbllx=31pt,bblly=145pt,bburx=545pt,bbury=671pt,clip=}}
\caption{\small $\alpha_s$ as a function of energy below
20 GeV.
The solid line represents  the non-perturbative $\alpha_s(E)$
as a function of energy  $E$ of order $m$
calculated from  Eq. (\ref{sof}).
The dotted lines are defining the $\pm$1 s.d. band.
The curved dashed line labeled as pQCD is the perturbative $\alpha_s$
calculated using reference \cite{bethker} with
the number of flavors N$_f$=3 and setting
$\Lambda_{\overline{M}\overline{S}}=200$ MeV
}
\label{nqcd}
\end{figure}

Our main results are shown in Fig. \ref{nqcd}. Since our system
in the center of mass energy is non-relativistic,
the $\alpha_s$ that we determine corresponds to an energy scale of $E\sim m$.
\noindent
The non-perturbative $\alpha_s$ is calculated via Eq. (\ref{sof})
and is shown in Fig. \ref{nqcd} as a function of energy by the solid line.
The accompanying dotted lines
represent the $\pm$1 s.d. limits of the  band.
For comparison we also show
the perturbative $\alpha_s$ curve, labeled by pQCD,
which essentially overlaps with the non-perturbative strong
coupling in the region of about 2 to 11 GeV. Using our
low energy non-perturbative strong coupling we obtain for
example $\alpha_s$ at the mass energy of the K-meson and the
$\Lambda$-hyperon respectively the following values:
$$\alpha_s(0.494\ {\rm{GeV}})=0.451\pm 0.035\ \ \ \ {\rm{and}}\ \ \ \
\alpha_s(1.115\ {\rm{GeV}})=0.392\pm 0.019$$\ .
The non-perturbative $\alpha_s$
determined here as a
function of energy
agrees well with the results obtained in \cite{salvi1,salvi2}, where effective $\alpha_s$ was
obtained by solving the Bethe-Saltpeter equation for quarkonium.
As a result the strong coupling constant $\alpha_s$ determined here agrees well
with the  one-loop  Analytic Perturbative Theory
(APT) approach \cite{salvi3}).  In particular we have  good agreement with the so called
"massive" variation of the APT prescription \cite{salvi3,salvi4}.
The latter one approach coincide with the standard APT approach for
energy scales of $E>200$ MeV \cite{shirkov}, i.e. above the pion mass.
However for small $E$ the strong coupling constant goes down
to zero\cite{salvi3,salvi4}, exactly as we have in
Fig. \ref{nqcd}.
It is worthwhile to note however, that although our curve is in good
agreement with the "massive" APT prescrition at small $E$
of order the pion mass, strictly speaking it is questionable if we can
apply
at the pion scale our estimates, based on simple nonrelativistic
quark model. Thus "reasonable" agreement of our result for the pion
mass with a particular version of APT prescription deserves further study.
\vspace{2mm}
\par Note that recently additional approaches to the determination of $\alpha_s$ behaviour in the low transverse momenta region were 
discussed, see \cite{srigo},\cite{cvetic}. The approach in \cite{srigo} is based on one-loop renormalization loop calculations, while in 
\cite{cvetic} on combining dispersion relations with lattice simulations results and experoimental data on $e^+e^-$ annihilation.
It is quite amazing that all the approaches lead to the same qualitative form of $\alpha_s$ as a function of energy/transverse 
momentum scale.
\par On the other hand, it is clear that all theoretical descriptions of the infrared dynamics at the moment are model dependent and based,
from field theoretical point of view , on different ansats for field theoretical resummation of corrections. 
In particular there arises a question of connection between strong coupling constant in nonrelativistic quark model used in this paper and 
$\alpha_s$ defined in field theoretical schemes for infrared QCD dynamics. This question was discussed in details in \cite{salvi1,salvi2,salvi3,salvi4}
where it was shown that $\alpha_s$ defined in particular renormalization/resummation scheme in these references, can indeed be identified
with (up to short range corrections) with the running coupling constant that enters the potential for nonrelativistic bound states.
\noindent
\par In conclusion the strong coupling constant $\alpha_s(E)$ can be
evaluated in the non-perturbative
region by the use of the Bose-Einstein and Fermi-Dirac correlations dimension results.
The resulting $\alpha_s(E)$ is in good agreement with the so called
APT "massive"  prescription \cite{salvi3,salvi4}
and extrapolates well at the higher energies
to the conventional perturbative $\alpha_s(E)$.
Our results indicate that
the BEC/FDC correlations both for baryons and  mesons,
correspond to a picture
where the two participating hadrons strongly overlap, and the \RI radius,
that conventionally characterizes the scale of the BEC/FDC, corresponds
to the distance $r$ between the centers of these two
correlated particles. Thus our results indicate that these correlations
may well serve as an alternative
approach for the  study of short range correlations between hadrons \cite{FS}.
\vspace{2mm}

\noindent
{\bf Acknowledgments:} One of us (B. Blok) would like to thank M. Strikman
for useful discussions.

%\end{flushleft}
\end{document}